\documentclass{article}
\topskip=\baselineskip
\let\al=\alpha
\let\bt=\beta
\let\gm=\gamma
\let\dl=\delta
\let\ep=\epsilon

\let\sg=\sigma
\let\la=\langle
\let\ra=\rangle
\let\pa=\partial
\let\e=\emph

\let\ct=\cite

\let\bv=\mathbf
\let\mr=\mathrm
\let\dt=\cdot
\let\del=\nabla
\let\dg=\dagger

\let\q=\widehat

\let\h=\hbar

\let\rta=\rightarrow
\let\lra=\leftrightarrow
\let\x=\times
\let\dy=\displaystyle
\let\ty=\textstyle

\let\hl=\hfill
\newcommand{\m}{\mbox}
\newcommand{\ol}[1]{\makebox[\textwidth][s]{#1}}
\newcommand{\id}{\mathrm{I}}
\newcommand{\eqdf}{\stackrel{\mathrm{def}}{=}}
\newcommand{\hf}{\ensuremath{{\scriptstyle\frac{1}{2}}}}
\newcommand{\hfs}{\ensuremath{{\scriptscriptstyle\frac{1}{2}}}}

\newcommand{\be}{\begin{equation}}
\newcommand{\ee}{\end{equation}}
\newcommand{\dd}[3]{\\ \m{}\\ \ol{\m{#1}\hl\m{${\dy #2}$}\hl\m{#3}}\\ \m{}\\}
\newcommand{\re}[2]{\dd{}{#1}{(#2)}}

\newcommand{\ba}{\begin{array}}
\newcommand{\ea}{\end{array}}
\newcommand{\bea}{\begin{eqnarray}}
\newcommand{\eea}{\end{eqnarray}}
\newcommand{\beas}{\begin{eqnarray*}}
\newcommand{\eeas}{\end{eqnarray*}}

\newcommand{\qH}{\q{H}}
\newcommand{\qHF}{\q{H}_F}

\newcommand{\Hfr}{H_{\mr{free}}}
\newcommand{\qHfr}{\q{H}_{\mr{free}}}

\newcommand{\Hnz}{H^{\mr{(NR)}}_{\mr{EM;0}}}
\newcommand{\Hrz}{H^{\mr{(REL)}}_{\mr{EM;0}}}

\newcommand{\Hnh}{H^{\mr{(NR)}}_{\mr{EM;}\hfs}}
\newcommand{\Hrh}{H^{\mr{(REL)}}_{\mr{EM;}\hfs}}
\newcommand{\qHrh}{\q{H}^{\mr{(REL)}}_{\hfs}}
\newcommand{\psmi}{(\psi_-(\vr))_i}
\newcommand{\psmdi}{(\psi_-^\dg(\vr))_i}
\newcommand{\pspi}{(\psi_+(\vr))_i}
\newcommand{\pspdi}{(\psi_+^\dg(\vr))_i}
\newcommand{\psmj}{(\psi_-(\vr'))_j}
\newcommand{\psmdj}{(\psi_-^\dg(\vr'))_j}
\newcommand{\pspj}{(\psi_+(\vr'))_j}
\newcommand{\pspdj}{(\psi_+^\dg(\vr'))_j}
\newcommand{\isis}{{\ty\int d^3\vr\sum_{i=1}^2\int d^3\vr'\sum_{j=1}^2}}

\newcommand{\vr}{\bv{r}}
\newcommand{\vv}{\bv{v}}

\newcommand{\vp}{\bv{p}}
\newcommand{\vP}{\bv{P}}

\newcommand{\vA}{\bv{A}}

\newcommand{\vL}{\bv{L}}

\newcommand{\vz}{\bv{0}}

\newcommand{\qvr}{\q{\vr}}

\newcommand{\qvp}{\q{\vp}}
\newcommand{\qvP}{\q{\vP}}
\newcommand{\qvL}{\q{\vL}}

\title{Do experiment and the correspondence principle \\
       oblige revision of relativistic quantum theory?}
\author{Steven Kenneth Kauffmann \\
        American Physical Society Senior Life Member}
\date{43 Bedok Road \\
      {\#}01-11 \\
      Country Park Condominium \\
      Singapore 469564 \\
      Handphone: +65 9370 6583 \\
      \m{} \\
      and \\
      \m{} \\
      Unit 802, Reflection on the Sea \\
      120 Marine Parade \\
      Coolangatta QLD 4225 \\
      Australia \\
      Tel/FAX: +61 7 5536 7235 \\
      Mobile:  +61 4 0567 9058 \\
      \m{} \\
      Email: SKKauffmann@gmail.com}
\begin{document}
\maketitle
\begin{abstract}
Recent preliminary data gathered by the Fermilab MINOS Collaboration suggest
with 95\% confidence that the mass of the muon neutrino differs from that of
its antineutrino partner, which contradicts the entrenched relativistic quantum
theory notion that a free antiparticle is a negative-energy free particle
compelled to travel backwards in time.   Also a discrepancy of about five
standard deviations in the value of the proton charge radius recently obtained
from muonic hydrogen versus that previously obtained from electronic hydrogen
casts doubt on the calculation of the dominant relativistic QED contributions
to the effects that are actually measured (e.g., the Lamb shift): these QED
contributions dominate proton charge radius contributions less in muonic
hydrogen than in electronic hydrogen.  The negative-energy ``free particles''
of entrenched relativistic quantum theory are well-known features of the
Klein-Gordon and Dirac equations, which are shown to have many other unphysical
features as well.  The correspondence principle for relativistic particles is
incompatible with these two equations, produces no unphysical features and
implies only positive energies for free particles, which eliminates the very
basis of the entrenched notion of antiparticles, as well as of the CPT theorem.
This principle thus requires antiparticles to arise from charge conjugation (or
more generally CP) invariance, whose known breaking is naturally expected to
produce mass splitting between particle and antiparticle, in consonance with
the preliminary MINOS data.  It also requires revamping of relativistic QED,
which is in accord with the doubt cast on it by the proton charge radius
results, and implies that QED is nonlocal, i.e. has no Hamiltonian density.
\end{abstract}

\subsection*{Introduction}

Recent data gathered by two very different experiments have cast a
shadow of doubt over the validity of relativistic quantum precepts
that have become well-entrenched over almost nine decades.  Prelim%
inary data from the Fermilab MINOS Collaboration presented on June
14, 2010 at the Neutrino 2010 conference in Athens, Greece suggest
with 95\% confidence that the muon neutrino does not have the same
mass as the muon antineutrino~\ct{Vh, Cw}.  If the symmetry which
relates particle to antiparticle were deemed to be a multi-particle
one of the overlying field theory, as is the symmetry which relates
the two members of an isospin doublet, such a mass splitting be%
tween neutrino and antineutrino would be \e{no more remarkable}
than is the mass splitting between proton and neutron: after all,
just as electromagnetism \e{breaks} isospin symmetry, there is a
physical agency which \e{breaks} particle-antiparticle symmetry---%
that is clear from particle domination of the composition of the
visible universe.

The issue, however, is that the entrenched approach to relativistic
quantum theory has it that the relation between particle and anti%
particle is \e{not} a mere multi-particle symmetry of the overlying
field theory, but that particle and antiparticle are in fact \e{two
members of the very same species}: a free antiparticle is deemed by
entrenched theory to be a free \e{negative}-energy particle which,
due to its negative energy, is somehow obliged to travel backwards
in time---although no deduction from established physics which jus%
tifies this astounding contention of time-flow reversal for nega%
tive-energy free particles is proffered.  This particular (and cer%
tainly peculiar) ``species identity'' of particle with antiparticle
in entrenched theory precludes their masses from differing \e{at
all}, and it as well lies at the \e{very heart} of the ``celebra%
ted'' CPT theorem.

The above-noted \e{negative}-energy free particles of course
\e{arise} from the ostensible ``quantum relativistic'' Klein-Gordon
and Dirac equations of entrenched theory.  These negative energies
\e{have no lower bound}, and therefore at first glance comprise a
source of severe theoretical physics embarrassment for the Klein-%
Gordon and Dirac free-particle equations---not to mention that free
particles of negative energy are not observed.  Putting the Klein-%
Gordon and Dirac negative-energy free particles ``at the service''
of a phenomenon that actually \e{is} observed, namely antiparti%
cles, by arbitrarily imposing on them the mind-boggling requirement
that they \e{also} travel backwards in time turned out to be asto%
nishingly well-received by the physics community.  This almost cer%
tainly was due to great reluctance on the part of this community to
\e{discard} the Klein-Gordon and Dirac equations, notwithstanding
that the negative free-particle energies are \e{just one} of a
\e{list} of egregiously unphysical properties which these equations
possess~\ct{Ka0}: the Klein-Gordon and Dirac equations have a de%
cided attraction for working theorists because they tend to be
\e{very tractable in calculations}, in part because they are purely
\e{local} in configuration representation.

Indeed fondness for the calculational tractability of the Klein-%
Gordon and Dirac equations acted as a strong distraction from even
\e{awareness} of the \e{basic requirement} which the \e{correspond%
ence principle} imposes on the quantum theory of relativistic free
particles, namely that the \e{classical} relativistic Hamiltonian
for a free particle of positive mass $m$,
\re{
\Hfr = (m^2c^4 + |c\vp|^2)^\hf,
}{1a}
is to be \e{quantized}, upon which it straightforwardly becomes the
\e{positive-definite} free-particle relativistic Hamiltonian
\e{operator},
\re{
\qHfr = (m^2c^4 + |c\qvp|^2)^\hf,
}{1b}
which is, of course, the essential input to the time-dependent
Schr\"{o}dinger equation for the relativistic free particle of pos%
itive mass $m$,
\re{
i\h\pa(|\psi(t)\ra)/\pa t =
                (m^2c^4 + |c\qvp|^2)^\hf|\psi(t)\ra.
}{1c}
\indent
\e{Heeding} the correspondence principle for relativistic free par%
ticles therefore \e{requires} that the time-dependent relativistic
free-particle Schr\"{o}dinger equation of Eq.~(1c) \e{must sup%
plant} the ``more tractable'' free-particle Klein-Gordon equation.
It \e{even} must supplant the free-particle Dirac equation: the
\e{nonrelativistic} Pauli equation for the spin~$\hf$ particle has
\e{no spin dependence whatsoever} when that particle is \e{free},
and furthermore there \e{always} exists an inertial frame in which
a free particle moves \e{nonrelativistically} (or is even at rest).
It is of course clear that the relativistic free-particle Hamilton%
ian operator $\qHfr$ of Eq.~(1b) \e{has no negative eigenenergies
whatsoever}.  Thus \e{enforcement} of the correspondence principle
\e{automatically} makes it \e{impossible} to even \e{speak about}
the mind-boggling notion of ``negative-energy free particles that
travel backwards in time'', which \e{forecloses} any possibility of
characterizing antiparticles as such, eliminating the basis of the
CPT theorem and its corollaries.

In the context of \e{respecting} the correspondence principle, an%
tiparticles obviously must be introduced at the multi-particle le%
vel via the imposition of CP invariance on the field-theory Hamil%
tonian---in the longer run the nature of the physical mechanisms
which in fact \e{break} CP invariance will need to be discovered.
Of course in this context of \e{respecting} the correspondence
principle there is no reason at all why these CP-breaking mechan%
isms should not produce particle-antiparticle mass splitting in
consonance with what the preliminary data from MINOS appear to
indicate.

The relativistic free-particle Klein-Gordon and Dirac equations,
notwithstanding their associated lists of unphysical features%
~\ct{Ka0}, have, of course, been clung to by those who do calcula%
tions partly because they are \e{local} in configuration represen%
tation, which inter alia results in \e{local} field theories.  The
relativistic free-particle time-dependent Schr\"{o}dinger equation
of Eq.~(1c), which follows from the correspondence principle,
has \e{no} unphysical features that correspond to either those of
the Klein-Gordon or the Dirac equation~\ct{Ka0}, but its representa%
tion in configuration space is \e{nonlocal}, so consequent quantum
field theories will \e{as well} be formally nonlocal, i.e., their
field-theory Hamiltonians will \e{not} have underlying \e{Hamilton%
ian densities} in the configuration regime.

Indeed there are \e{other} details of specifically the \e{quantum
electrodynamics} which the correspondence principle implies that
\e{must differ} from those of the present theory, in which the Di%
rac equation figures so prominently.  In particular, the Dirac
equation in the presence of an external electromagnetic field needs
to be \e{replaced} by a time-dependent Schr\"{o}dinger equation
which smoothly \e{reduces} to the time-dependent relativistic
\e{free-particle} Schr\"{o}dinger equation that is given by
Eq.~(1c) when that external electromagnetic field is switched off.
That time-dependent Schr\"{o}dinger equation must \e{also} smoothly
reduce to the nonrelativistic Pauli equation in the nonrelativistic
limit.  Such an equation has indeed been developed from the nonrel%
ativistic Pauli equation by systematically applying to it fully
relativistic upgrading techniques which are guided by the basic ob%
servation that there always exists an inertial frame in which a
positive-mass solitary particle is instantaneously moving nonrela%
tivistically~\ct{Ka0}.  As pointed out above, antiparticles must be
brought into correspondence-principle compatible quantum electrody%
namics by imposing charge-conjugation invariance on the field-theo%
ry Hamiltonian (parity, of course, is \e{conserved} in electrody%
namics).  After this is done, particle-antiparticle pair production
and annihilation is made possible by the imposition on the field-%
theory Hamiltonian of a \e{further} symmetry, namely its invariance
under the interchange of particle annihilation with antiparticle
creation as well as under the interchange of antiparticle annihila%
tion with particle creation.

Of course the correspondence-principle compatible quantum electro%
dynamics of a relativistic spin 0 charged particle of positive mass
is to be handled in closely similar fashion; there the analogous
systematic relativistic upgrade of the nonrelativistic Hamiltonian
of a spinless, positive-mass charged particle in completely nonrel%
ativistic interaction with an electric potential neatly results in
precisely the fully relativistic Hamiltonian from which Hamilton's
classical equations of motion produce the fully relativistic ver%
sion of the Lorentz force law~\ct{Ka0}.  This brings to light sub%
tle and important physics that the Klein-Gordon equation, which in%
herently reflects only the \e{square} of a Hamiltonian~\ct{Ka0}, is
obviously incapable of fully emulating.

Elucidation of the \e{full} structure of the modified quantum electro%
dynamics that is rooted in the requirements of the correspondence
principle, right up to and including its ``Feynman rules'', requires a
quite massive investment of time, patience, and ingenuity on the part
of a host of contributors.  It is furthermore naturally to be expected
that the predictions of the modified theory will \e{deviate} somewhat
from the predictions of the currently existing quantum electrodynamics
in which the physically problematic Dirac or Klein-Gordon equations
figure so prominently.

It is very interesting is this regard that a recent effort to ob%
tain the value of the charge radius of the proton to high precision
from measurement of the Lamb shift in muonic hydrogen has produced
a result which is incompatible with the value of this charge radius
that is obtained from combining precision spectroscopy of electron%
ic hydrogen with bound-state quantum electrodynamics~\ct{Po}. The
Lamb shift itself is, of course, overwhelmingly due to a bound-%
state quantum electrodynamics effect (it vanishes in the nonrelati%
vistic Schr\"{o}dinger and in the relativistic Dirac equation mo%
dels of the hydrogen atom), with only a very small percentage con%
tribution to it arising from the charge radius of the proton, al%
beit that very small proton charge radius percentage contribution
is clearly very much greater (up to 2\%~\ct{Po}) for muonic hydro%
gen than it would be for electronic hydrogen, whose Bohr radius is
about two hundred times larger.  Generally speaking, this very big
Bohr radius difference implies that the importance of quantum elec%
trodynamics calculations for the extraction of the charge radius of
the proton from hydrogen atomic spectroscopy looms very much larger
for electronic than for muonic hydrogen, notwithstanding that it is
already very important for the latter.  The above-mentioned two in%
compatible results (about five standard deviations discrepancy%
~\ct{Po}) for the proton charge radius naturally casts suspicion on
the present theoretical form of quantum electrodynamics in light of
the far larger contribution made by quantum electrodynamics than by
the proton charge radius itself to the effects that are actually
measured---\e{especially} in view of the fact that the quantum
electrodynamics contributions are systematically even \e{more}
dominant over the proton charge radius contribution in electronic
hydrogen than they are in muonic hydrogen.

In the following sections key theoretical physics issues alluded to
in the preceding paragraphs are treated at length along the lines
expounded in Ref.~\ct{Ka0}.  We begin by pointing out the \e{natural
compatibility} of solitary-particle quantum mechanics with special
relativity, which consequently \e{reaffirms the validity} of the
\e{correspondence principle} in the domain of solitary-particle rela%
tivistic quantum mechanics, and we \e{also} point out the reason why
\e{only} the Hamiltonian of Eq.~(1a) is suitable for a relativistic
classical free particle of positive mass $m$.

\subsection*{Solitary-particle quantum mechanics' inherent compati%
             bility with relativity}

The compatibility of solitary-particle quantum mechanics with spe%
cial relativity is a straightforward consequence Schr\"{o}dinger's
two basic postulates for the wave function~\ct{Scf,B-D}, namely
$\la\vr|\psi(t)\ra$.  The \e{first} Schr\"{o}dinger postulate is the
wave-function rule for the \e{operator quantization} of the parti%
cle's \e{canonical three-momentum},
\re{
    \la\vr|\qvp|\psi(t)\ra = -i\h\del_{\vr}(\la\vr|\psi(t)\ra),
}{2a}
which is as well, of course, a result of Dirac's postulated
canonical commutation rule~\ct{Dir}.

The \e{second} Schr\"{o}dinger wave-function postulate is the famed
\e{time-dependent Schr\"{o}dinger wave equation}~\ct{Scf,Dir,B-D},
\re{
    i\h\partial(\la\vr|\psi(t)\ra)/\partial t = \la\vr|\qH|\psi(t)\ra,
}{2b}
which treats the \e{operator quantization} $\qH$ of the particle's
classical Hamiltonian $H$ in a manner that is \e{formally parallel} to
the way in which Eq.~(2a) treats the \e{operator quantization} of the
particle's canonical three-momentum.  The straightforward theoretical
physics \e{implication} of Eqs.~(2a) and (2b) is simply that the oper%
ators $\qvp$ and $\qH$ are the \e{generators} of the wave function's
\e{infinitesimal space and time translations}, respectively.  There%
fore, in anticipation of the \e{restriction on such generators which
special relativity imposes}, these two equations are usefully \e{com%
bined} into the \e{single} formally \e{four-vector} Schr\"{o}dinger
equation for the wave function,
\re{
    i\h\partial(\la\vr|\psi(t)\ra)/\partial x_{\mu} =
                \la\vr|\q{p^{\mu}}|\psi(t)\ra,
}{2c}
where the \e{contravariant four-vector space-time partial derivative
operator} $\pa/\pa x_{\mu}$ is defined as
$\pa/\pa x_{\mu}\eqdf(c^{-1}\pa/\pa t, -\del_{\vr})$, and the \e{formal}
``contravariant four-vector'' energy-momentum operator $\q{p^{\mu}}$ is
defined as $\q{p^{\mu}}\eqdf(\qH/c, \qvp)$.  Since special relativity 
\e{requires} the contravariant space-time partial derivative four-vector
operator $\pa/\pa x_{\mu}$ to transform between inertial frames in
\e{Lorentz-covariant} fashion, it is apparent from Eq.~(2c) that the
\e{Hamiltonian operator} $\qH$ will be \e{compatible with special
relativity} if it is related to the canonical three-momentum operator
$\qvp$ in such a way that \e{also} makes the energy-momentum operator
$\q{p^{\mu}}$ a contravariant four-vector which transforms between
inertial frames in \e{Lorentz-covariant} fashion.  This property of the
Hamiltonian operator will, of course, be \e{automatically satisfied} if
it is \e{the quantization of the Hamiltonian of a properly relativistic
classical theory}.  Therefore the \e{correspondence principle} defin%
itely \e{remains valid} in the solitary-particle special-relativistic
domain!

Now for a relativistic classical free particle of positive mass $m$,
the logic of the Lorentz transformation from its \e{rest frame}, where
it has four-momentum $(mc, \vz)$, to a frame where it has velocity
$\vv$(where $|\vv| < c$) leaves \e{no freedom at all in the choice of
its classical Hamiltonian}.  That Lorentz boost takes this particle's
four-momentum to,
\re{
    (mc(1-|\vv|^2/c^2)^{-\hf},\: m\vv(1-|\vv|^2/c^2)^{-\hf}) =
    (E(\vv)/c,\: \vp(\vv)),
}{3a}
which, together with the \e{identity},
\re{
    mc^2(1-|\vv|^2/c^2)^{-\hf} =
    \sqrt{m^2c^4 + |cm\vv|^2(1 -|\vv|^2/c^2)^{-1}},
}{3b}
implies that,
\re{
E(\vv) = \sqrt{m^2c^4 + |c\vp(\vv)|^2} = \Hfr(\vp(\vv)).
}{3c}
Therefore the \e{only} physically suitable Hamiltonian for the relati%
vistic classical free particle of positive mass $m$ is the $\Hfr$ of
Eq.~(1a).  Thus \e{adherence to the correspondence principle}, togeth%
er with the \e{categorical implication} of Eqs.~(3), \e{determines}
the Hamiltonian operator for the relativistic free particle of posi%
tive mass $m$ to be the square-root operator given by Eq.~(1b),
namely,
\[ \qHfr = \sqrt{m^2c^4 + |c\qvp|^2},\]
which implies that the time-dependent Schr\"{o}dinger equation for the
relativistic free particle of positive mass $m$ is that of Eq.~(1c).

Since Eq.~(1c) is therefore the \e{only quantum physically correct}
time-dependent description of the relativistic free particle of posi%
tive mass $m$, the free-particle Klein-Gordon and Dirac equations ipso
facto \e{must be quantum physically defective}.  We now proceed to
\e{analyze the sources} of those physical defects and also to \e{list}
some of the \e{unphysical consequences} of the free-particle Klein-%
Gordon and Dirac equations.

\subsubsection*{The physically unsuitable Hamiltonian-squared basis of
                the free-particle Klein-Gordon equation}

Because the square-root Hamiltonian operator $\qHfr$ of Eq.~(1b) for
the positive-mass relativistic free particle is nonlocal in configura%
tion representation, which might conceivably present an awkward calcu%
lational hurdle at a later stage when interactions with an external
field are added, Klein, Gordon and Schr\"{o}dinger \e{rejected} the
\e{physically correct} positive-mass relativistic free-particle time-%
dependent Schr\"{o}dinger equation of Eq.~(1c) in favor of its \e{it%
eration}, which \e{squares} its square-root Hamiltonian operator
$\qHfr$, and, in conjunction with Schr\"{o}dinger's canonical three-%
momentum quantization rule of Eq.~(2a), yields,
\re{
    -\h^2\pa^2(\la\vr|\psi(t)\ra)/\pa t^2 =
    (m^2c^4 - \h^2c^2\del^2_{\vr})\la\vr|\psi(t)\ra,
}{4a}
which is readily rewritten in the customary form for the free-particle
Klein-Gordon equation,
\re{
    (\pa^2/(\pa x^{\mu}\pa x_{\mu}) + (mc/\h)^2)\la\vr|\psi(t)\ra = 0.
}{4b}
To each stationary eigensolution $e^{-i\sqrt{m^2c^4 +
|c\vp |^2}\:t/\h}\la\vr|\vp\ra$ of eigenmomentum $\vp$ of the physi%
cally correct time-dependent relativistic free-particle Schr\"{o}din%
ger equation, given by Eq.~(1c), Eq.~(4a) \e{adds} an \e{extraneous
negative-energy partner solution} $e^{+i\sqrt{m^2c^4 +
|c\vp |^2}\:t/\h}\la\vr|\vp\ra$ of the \e{same} momentum, whose
\e{sole reason for existing} is the \e{entirely gratuitous
iteration} of Eq.~(1c)!  These \e{completely} extraneous \e{negative}
``free solitary-particle'' energies, $-\sqrt{m^2c^4 + |c\vp |^2}$, do
\e{not} correspond to \e{anything} that exists in the \e{classical}
dynamics of a free relativistic solitary particle, and by their nega%
tively \e{unbounded} character threaten to spawn unstable runaway
phenomena should the \e{free} Klein-Gordon equation be sufficiently
perturbed (the Klein paradox)~\ct{B-D}.

Due to the fact that the free-particle Klein-Gordon equation \e{lacks}
a corresponding Hamiltonian operator---it depends on \e{only} the
\e{square} of the Hamiltonian operator $\qHfr$, as is seen from
Eq.~(4a) in conjunction with Eq.~(2a)---it turns out, as is easily
verified, that the \e{two} solutions of the \e{same momentum} $\vp$
which have \e{opposite-sign} energies, i.e., $\pm\sqrt{m^2c^4 +
|c\vp|^2}$, \e{fail to be orthogonal to each other}, which \e{out%
right violates a key property} of orthodox quantum mechanics! 
\e{Without this property} the probablity interpretation of quantum
mechanics \e{cannot be sustained}, and the Klein-Gordon equation is
unsurprisingly \e{diseased} in that regard, yielding, inter alia,
\e{negative probabilities}~\ct{B-D}.  

Furthermore, free-particle Klein-Gordon theory, depending as it does
on \e{only} the \e{square} of the Hamiltonian operator $\qHfr$ of
Eq.~(1b), rather than on that Hamiltonian operator \e{itself}, is
thereby \e{cut adrift} from the normal quantum mechanical relationship
to the Heisenberg picture, Heisenberg's equations of motion and the
Ehrenfest theorem.

The fact of the matter is that the \e{square} of a Hamiltonian opera%
tor, \e{unlike} that Hamiltonian operator \e{itelf}, has \e{no cogent
physical meaning}!  That is the \e{source} of the above list of \e{un%
physical consequences} of the free-particle Klein-Gordon equation.

\subsubsection*{Space-time mishandling of Schr\"{o}dinger's equa%
                tion that engenders Dirac's free-particle equation}

Dirac pondered the foregoing list of the free-particle Klein-Gordon
equation's unphysical properties, especially its failure to have a
probability interpretation, and concluded that its dependence on
\e{only} the \e{square} of the Hamiltonian operator $\qHfr$ of Eq.~%
(1b) was not tenable, but that the time-dependent description of a
quantum mechanical system instead \e{must} be couched in terms of a
time-dependent Schr\"{o}dinder equation of the form of Eq.~(2b) with a
Hermitian Hamiltonian operator $\qH$.  Very unfortunately indeed, not%
withstanding that the correspondence principle \e{mandates} that this
$\qH$ \e{must} equal the $\qHfr$ of Eq.~(1b) for the case of a posi%
tive-mass relativistic free particle, Dirac, emulating Klein, Gordon
and Schr\"{o}dinger, continued to \e{reject} the \e{physically cor%
rect} square-root Hamiltonian operator $\qHfr$ of Eq.~(1b) for the
positive-mass relativistic free particle out of concern that its non%
locality in configuration representation might present an awkward cal%
culational hurdle at a later stage when interactions of that particle
with an external field are included.

Casting about for a more compelling theoretical ``justification'' than
mere concerns over conceivable calculational hurdles for his quantum-%
physically \e{untenable} rejection of the square-root Hamiltonian op%
erator $\qHfr$, Dirac hit upon a spurious ``relativistic need'' for
the time-dependent Schr\"{o}dinger equation of Eq.~(2b) to \e{by it%
self} exhibit ``space-time coordinate symmetry''~\ct{D28, Scw, B-D}.

It is, of course, abundantly clear that it is the \e{four-vector}
Schr\"{o}dinger equation \e{system} of Eq.~(2c) which \e{in fact mani%
fests just this space-time coordinate symmetry} when its Hamiltonian
operator $\qH$ is related to the canonical three-momentum operator
$\qvp$ in such a way that the energy-momentum operator $\q{p^{\mu}} =
(\qH/c, \qvp)$ is a contravariant four-vector which transforms between
inertial frames in \e{Lorentz-covariant} fashion, a property of $\qH$
which is \e{automatically satisfied} when it is \e{the quantization of
a Hamiltonian} $H$ \e{of a properly relativistic classical theory}!
Thus the time-dependent Schr\"{o}dinger equation of Eq.~(2b) upon
which Dirac myopically fastened his ``space-time coordinate symmetry''
gaze is the mere \e{time component} of a Lorentz-covariant \e{four-%
vector equation system}, and, as such, is \e{not space-time coordin%
ate symmetric at all} since it is \e{completely skewed toward time}!

The fact of this \e{utter skewing toward time} of the time-dependent
Schr\"{o}dinger equation of Eq.~(2b) is \e{driven home} by the \e{the%
oretical physics content} which its mathematical presentation \e{un%
mistakably conveys}, namely that the Hamiltonian operator is the gen%
erator of the \e{time translations} of the wave function.  To attempt
to \e{force} ``space-time coordinate symmetry" on an equation which is
\e{so completely skewed toward time} as is the time-dependent
Schr\"{o}dinger equation is a \e{classic} instance of attempting to
``jam a square peg into a round hole'', and can \e{only} result in a
plenitude of unphysical consequences.

Blithely insensitive to the \e{necessarily} completely time-skewed na%
ture of the time-dependent Schr\"{o}dinger equation of Eq.~(2b), Dirac
noted that its \e{left-hand side} is \e{proportional} the \e{time-der%
ivative} operator $\pa/\pa t$, and therefore sought to impose space-%
time coordinate symmetry on it by requiring its \e{right-hand side} to
be (inhomogeneously) \e{linear} in the \e{spatial gradient} operator
$\del_{\vr}$.  Of course the right-hand side of the time-dependent
Schr\"{o}dinger equation of Eq.~(2b) only involves the \e{Hamiltonian
operator $\qH$ in configuration representation}, whose (inhomogeneous)
\e{linearity} in $\del_{\vr}$ guarantees its \e{local nature}, which
of course was Dirac's \e{overriding consideration from the very begin%
ning}!

More abstractly, Dirac's imposition of space-time coordinate symmetry
on the configuration-representation time-dependent Schr\"{o}dinger
equation of Eq.~(2b) implies that its Hamiltonian operator $\qH$ is
(inhomogeneously) \e{linear} in the \e{momentum operator} $\qvp$.  If
we now calculate the particle \e{velocity operator} $d\qvr/dt$ that is
\e{implied} by such a Hamiltonian operator, i.e., one  which is
\e{linear} in the momentum operator $\qvp$, by using Heisenberg's
equation of motion, we immediately obtain that this \e{velocity opera%
tor} $d\qvr/dt$ is \e{completely independent} of the \e{momentum oper%
ator} $\qvp$.  However, we know very well that for the postive-mass
relativistic \e{free particle in the nonrelativistic regime} the velo%
city operator $d\qvr/dt$ is \e{proportional} to $\qvp$ (i.e., equals
$\qvp/m$), and, more generally, the \e{relativistic free particle} is
\e{always} expected to have its velocity operator $d\qvr/dt$ \e{paral%
lel} to the momentum operator $\qvp$, but this is obviously \e{imposs%
ible} if $d\qvr/dt$ is \e{independent} of $\qvp$, which is the
\e{clear consequence} of Dirac's \e{physically misconceived} effort
to \e{force} space-time coordinate symmetry on the time-dependent
Schr\"{o}dinger equation of Eq.~(2b).  In \e{stark contrast}, if
we use for the Hamiltonian operator $\qH$ in the time-dependent
Schr\"{o}dinger equation of Eq.~(2b) the positive-mass relativistic
free-particle \e{square-root} Hamiltonian operator $\qHfr$ of Eq.~%
(1b) that is \e{mandated by the correspondence principle}, Heisen%
berg's equation of motion yields,
\[ d\qvr/dt = \qvp/(m^2 + |\qvp/c|^2)^\hf,\]
which is \e{obviously the correct result}!  In other words, the
\e{upshot} of the \e{squirming} by Klein, Gordon, Schr\"{o}dinger, and
Dirac to \e{evade the mandate of the correspondence principle} only
results in \e{gratuitous theoretical grief} in the \e{completely un%
necessary form of obviously unphysical results}.

It is clear that \e{continuing} with Dirac's physically misconceived
approach is \e{counterproductive} from the standpoint of attaining
physically correct understanding of positive-mass relativistic soli%
tary-particle quantum mechanics.  However, it is the case that text%
books~\ct{B-D, Scw, B-S} have simply \e{not presented} the most
\e{strikingly} unphysical consequences of Dirac's approach to the pos%
itive-mass relativistic free particle, which makes it worthwhile to
continue with Dirac's development in order to \e{expose those results
to the light of day}.

Dirac's physically misconceived imposition of space-time coordinate
symmetry on the time-dependent solitary-particle Schr\"{o}dinger equa%
tion of Eq.~(2b) does \e{not} fully determine its Hamiltonian operator
$\qH$; it \e{only} determines that $\qH$ is (inhomogeneously) \e{line%
ar} in the components of the momentum operator $\qvp$.  For the
\e{free} particle of positive mass $m$, we can write such a $\qH$ as,
\re{
\qH_D = \vec\al\dt\qvp c + \bt mc^2,
}{5a}
where what is known about $\bt$ and the components of $\vec\al$ is
that they are obviously \e{dimensionless}, and, because the solitary
particle is \e{free}, they \e{won't} depend on the particle's coordi%
nate operator $\qvr$, and so are \e{constants} in the particle's quan%
tized phase-space vector operator $(\qvr, \qvp)$.  Since that is
\e{all} that can be said about $\bt$ and $\vec\al$ \e{without} any
further assumption, Dirac decided to \e{make} an assumption which
essentially \e{determines} $\bt$ and $\vec\al$.  Having up to this
point \e{deliberately snubbed} the positive-mass relativistic free-
particle \e{square-root} Hamiltonian operator $\qHfr$ of Eq.~(1b)---%
which is \e{mandated} by the correspondence principle to in fact be
the \e{physically correct one}---Dirac now decided to \e{pull} $\qHfr$
\e{into the proceedings} by making it a \e{requirement} that,
\re{
    (\qH_D)^2 = (\qHfr)^2 = m^2c^4 + |c\qvp|^2.
}{5b}
Notwithstanding that this requirement \e{superficially} appears to be
a plausible one, Dirac \e{failed to note} that the \e{square} of a
Hamiltonian operator \e{has no cogent physical meaning}, just as
Klein, Gordon and Schr\"{o}dinger had \e{earlier failed to note this
very same pertinent fact}!  Setting equal two mathematical entities
which each lack definite physical meaning would seem \e{at least} as
likely to generate \e{unphysical} consequences as physically legiti%
mate ones.  Indeed the requirement of Eq.~(5b) turns out to be \e{di%
rectly responsible} for the fact that the eigenenergy spectrum of
$\qH_D$ \e{exactly matches} the energies of the solutions of the free%
-particle Klein-Gordon equation, \e{including} that equation's extran%
eous negative energies which are unbounded below!  So the \e{full}
theory of the free-particle Dirac Hamiltonian $\qH_D$ is underlain by
not merely \e{one}, but by \e{two} physically misconceived require%
ments.  It is perhaps little wonder, then, as we shall shortly see,
that $\qH_D$ gives rise to some stunningly unphysical predictions.

The upshot of the requirement of Eq.~(5b) turns out to be that $\bt$
and the three components of $\vec\al$ are Hermitian matrices (because
$\qH_D$ is required to be a Hermitian operator) which each square to
the identity matrix and which all mutually anticommute.  These proper%
ties of themselves imply that these four matrices are all as well
\e{traceless}~\ct{B-D}, which implies that $\qH_D$ is traceless as
well.  Therefore $\qH_D$ \e{must} have \e{negative} eigenvalues if it
has \e{positive} ones (and conversely).  This fact, taken together
with Eq.~(5b) itself, implies the aforementioned identity of the ei%
genenergy spectrum of $\qH_D$ with the energies of the solutions of
the free-particle Klein-Gordon equation, \e{including} that equation's
\e{extraneous negative energies}, which are \e{unbounded below}.

Returning now to the issue that was broached above concerning the free
Dirac particle's \e{velocity operator}, we obtain from Eq.~(5a) and
the Heisenberg equation of motion that,
\re{
d\qvr/dt = \vec\al c,
}{6a}
which has the \e{highly unphysical} property of being \e{completely
independent} of the particle's momentum operator $\qvp$, as was al%
ready pointed out above.  Even \e{worse}, the free particle's
\e{speed operator} comes out to be,
\re{
|d\qvr/dt| = \sqrt{3}\:c\id,
}{6b}
which stunningly has a \e{but a single eigenvalue that exceeds the
speed of light by} 73\%!  It is most interesting that while it is not
uncommon for textbooks to at least \e{mention} the velocity operator
result of Eq.~(6a)~\ct{B-D}---and then to rapidly turn away from it%
---there is apparently not a single textbook which \e{uses} Eq.~(6a)
to obtain the very simple consequent \e{speed} operator result of Eq.%
~(6b), which is, of course, utterly unphysical to an extent that is
breathtaking.  But underlain as the free-particle Dirac Hamiltonian
operator $\qH_D$ is by not merely \e{one} but actually \e{two} re%
quirements that are \e{not physically sensible}, namely the imposi%
tion of space-time symmetry on its time-dependent Schr\"{o}dinger
equation and the imposition on it of Eq.~(5b), it is perhaps not sur%
prising that it can give rise to such a blatantly relativistically-%
forbidden consequence.

Newton's first law of motion implies that the acceleration of a free
particle vanishes identically.  If we calculate $d^2\qvr/dt^2$ from
the positive-mass relativistic free-particle square-root Hamiltonian
operator $\qHfr$ of Eq.~(1b), which is mandated by the correspondence
principle, by applying Heisenberg's equation of motion twice in suc%
cession, we indeed obtain that this acceleration operator vanishes
identically.  It is a \e{very different story}, however, when we
switch this calculation to the free-particle Dirac Hamiltonian
$\qH_D$ of Eq.~(5a).  In that case, Heisenberg's equation of motion
yields,
\re{
 d^2\qvr/dt^2 = (imc^3/\h)(2\bt\vec\al + ((\vec\al\x\vec\al)\x\qvp)/(mc)),
}{7a}
which \e{fails} to vanish.  Note that the matrix cross product
$(\vec\al\x\vec\al)$ does \e{not} vanish because the three components
of $\vec\al$ mutually \e{anticommute}.  From Eq.~(7a) we can calculate
the magnitude of the free Dirac particle's spontaneous acceleration,
\re{
|d^2\qvr/dt^2| =  (2\sqrt{3}\:mc^3/\h)(1 + (2/3)(|\qvp|/(mc))^2)^\hf,
}{7b}
whose \e{minimum} value, $(2\sqrt{3}\:mc^3/\h)$, is, for the case of
the electron, well in excess of $10^{28} g$, where $g$ is the acceler%
ation of gravity at the earth's surface.  This dumbfounding spontan%
eous acceleration of the ``free Dirac electron'', which stupendously
violates Newton's first law of motion, again drives home the lesson of
\e{just how unphysical} the Dirac free-particle Hamiltonian $\qH_D$
is---but this result as well seems to have escaped the notice of
textbooks.

It is readily shown that the orbital angular momentum operator $\qvL
\eqdf\qvr\x\qvp$ commutes with the positive-mass relativistic free-%
particle square-root Hamiltonian operator $\qHfr$ of Eq.~(1b) that is
mandated by the correspondence principle.  It commutes as well with
the nonrelativistic free-particle Pauli Hamiltonian operator---which
is simply $|\qvp|/(2m)$ for that free-particle case.  However it does
\e{not} commute with the free-particle Dirac Hamiltonian $\qH_D$,
which yields the nonvanishing spontaneous spin-orbit torque operator,
\re{
    d\qvL/dt = \vec\al\x\qvp c,
}{8a}
whose magnitude is,
\re{
    |d\qvL/dt| = \sqrt{2}\:|\qvp|c,
}{8b}
Now the relativistic free particle's \e{kinetic energy} is,
\re{
\q{T} =  (m^2c^4 + |c\qvp|^2)^\hf - mc^2 = ((\qH_D)^2)^\hf - mc^2.
}{8c}
If we take the \e{dimensionless ratio} of the Dirac particle's spon%
taneous spin-orbit torque magnitude to its kinetic energy, we obtain,
\re{
 |d\qvL/dt|/\q{T} = \sqrt{2}\:((1 + (mc/|\qvp|)^2)^\hf + (mc/|\qvp|)),
}{8d}
which \e{increases monotonically without bound} from its ultrarelati%
vistic asymptotic value of $\sqrt{2}\:$ as $|\qvp|$ \e{decreases}.
This free-particle Dirac-theory result is, of course, \e{completely
inconsistent} with the free-particle Pauli theory, where this ratio
always \e{vanishes identically} for nonvanishing $|\qvp|$.

So the Dirac theory certainly \e{does not} reduce to the Pauli theory
merely by going to sufficiently small nonzero values of momentum.
That was \e{already clear}, of course, from the fact that the Dirac
particle's \e{speed} always has the value $\sqrt{3}\:c$ \e{irrespec%
tive} of its momentum, which \e{doesn't accord} with the free-parti%
cle Pauli theory speed operator $|\qvp|/m$ \e{at all} when $|\qvp|\ll
mc$.  The \e{highly anomalous} spontaneous spin-orbit coupling of the
\e{free} Dirac particle that we discussed above seems to have eluded
the notice of textbooks as well.

The examples of astoundingly unphysical results which emerge from the
Dirac free-particle Hamiltonian $\qH_D$ can apparently be multiplied
almost at will: e.g., the noncommutativity of orthogonal components of
the Dirac velocity operator of Eq.~(6a) has \e{surpassingly unphysi%
cal} systematic characterics,
\re{
[(d\qvr/dt)_x, (d\qvr/dt)_y] = 2c^2\al_x\al_y.
}{9}
This orthogonal velocity-component commutator \e{refuses to vanish}
even in the \e{classical limit} that $\h\rta 0$, in defiance of every%
thing known about classical velocity.  If one then struggles for a
glimmer of physical comprehension of this orthogonal velocity-compo%
nent commutator by going to the \e{nonrelativistic limit} $c\rta
\infty$, where it obviously \e{also} should vanish, it instead \e{di%
verges}!  The \e{highly unphysical behavior} of the \e{commutators} of
a list of \e{observables} in the free-particle Dirac theory has appar%
ently not been noticed by textbooks either.

\subsection*{Relativistic solitary-particle quantum mechanics in an
             electromagnetic potential}

The preceding subsections have made it abundantly clear that the
Klein-Gordon and Dirac theories cannot sensibly describe the positive%
-mass relativistic free particle, but the straightforward square-root
Hamiltonian operator $\qHfr$ of Eq.~(1b), \e{which is mandated by the
correspondence principle for this task}, describes the positive-mass
relativistic free particle flawlessly.  We shall now present in detail
the \e{extensions} of $\qHfr$ which were developed in Refs.~\ct{Ka1,
Ka0} for the cases of a solitary relativistic spin~0 and spin~$\hf$
particle of charge $e$ and positive mass $m$ in the presence of an
external electromagnetic potential $A^{\mu}(\vr, t)$.

The underlying idea is that if one has a trustworthy description of
the physics experienced by a solitary particle that moves \e{nonrelati%
vistically}, the physics that it experiences when it moves relativis%
tically boils down to Lorentz transformations from an appropriate suc%
cession of inertial frames in each of which it instantaneously moves
nonrelativistically.

However, instead of trying to model a self-consistently nonrelativis%
tic succession of inertial frames, and then carrying out the corres%
ponding Lorentz transformations, the technical approach adopted here
is rather to try to associate each individual term of the solitary
particle's \e{nonrelativistic Hamiltonian} with a fully Lorentz-covar%
iant four-momentum whose time component \e{reduces to that particular
nonrelativistic Hamiltonian term} in any inertial frame where the par%
ticle is moving sufficiently slowly.  All those individual Lorentz-co%
variant four-momenta are then \e{summed} to produce the solitary part%
icle's Lorentz-covariant \e{total} four-momentum.  The total three-mo%
mentum \e{part} of the solitary particle's total four-momentum is ob%
viously identified as the generator of the solitary particle's spatial
translations, and therefore as the solitary particle's relativistic
\e{canonical} three-momentum.  Of course the solitary particle's
\e{relativistic total energy}, when expressed as function of its rela%
tivistic \e{canonical} three-momentum, the time, and that particle's
three space coordinates comprises that particle's \e{relativistic Ham%
iltonian}.  \e{Initially}, of course, the \e{individual} four-momenta
that contribute to the solitary particle's \e{total} four-momentum
will be couched in the language of the particle's three space coordin%
ates, the time, and the particle's relativistic \e{kinetic} three-mom%
entum.  After \e{identification} of the particle's \e{canonical}
(i.e., \e{total}) three-momentum, it is \e{necessary} to solve for its
\e{kinetic} three-momentum as a \e{function} of its \e{canonical}
three-momentum in order to be able to \e{reexpress} its total energy
as its Hamiltonian.  Unfortunately, there is no guarantee that the
particle's relativistic \e{kinetic} three-momentum can be worked out
as a function of its relativistic \e{canonical} three-momentum \e{in
closed form}.  Thus the solitary particle's relativistic Hamiltonian
\e{itself} could \e{conceivably} only be available as a sequence of
approximations.

We begin by applying this program to a spin~0 solitary particle of
positive mass $m$ and charge $e$ in the presence of an external elec%
tromagnetic potential $A^\mu(\vr, t)$.  Note that all \e{magnetic} ef%
fects of such a potential on the spin~0 charged particle's motion
\e{vanish entirely} in the particle's rest frame, and are, more gener%
ally, of order $O(1/c)$, but in nonrelativistic physics the speed of
light $c$ is regarded as an asymptotically large parameter.  Thus the
\e{strictly nonrelativistic} Hamiltonian operator for this particle
involves \e{only} the electromagnetic potential's time component
$A^0(\vr, t)$,
\re{
    \q\Hnz = |\qvp|^2/(2m) + eA^0(\qvr, t).
}{10a}
Because of the technical issue regarding the choice of ordering of
noncommuting operators (whose resolution we allude to below), it is
convenient to develop the relativistic four-momentum as a function of
\e{classical} $(\vr, \vp)$ phase space \e{rather} than as a function of
the \e{already quantized} $(\qvr, \qvp)$ phase space of Eq.~(10a).  The
solitary particle's nonrelativistic kinetic energy $|\vp|^2/(2m)$, plus
its rest mass energy $mc^2$, is well-known to correspond to $c$ times
its Lorentz-covariant free-particle \e{kinetic}
four-momentum $p^\mu$,
\[p^\mu\eqdf((m^2c^2 + |\vp|^2)^\hf, \vp),\]
where, of course, $\vp$ is the particle's relativistic \e{kinetic}
three-momentum, which was carefully distinguished in the discussion
above from its relativistic \e{total} (i.e., \e{canonical}) three-%
momentum.  It is apparent that in the nonrelativistic limit
$|\vp|\ll mc$, the time component times $c$ of $p^\mu$ does indeed,
as just mentioned, behave as,
\[ cp^0 \approx mc^2 + |\vp|^2/(2m).\]
The potential energy term $eA^0(\vr, t)$ of $\Hnz$, divided by $c$, is
obviously the time component of the Lorentz-covariant four-momentum
$eA^\mu(\vr, t)/c$.  Therefore adding $eA^\mu/c$ to $p^\mu$ produces a
fully Lorentz-covariant \e{total} four-momentum whose time component
times $c$ reduces, in any inertial frame in which the nonzero-mass
charged spin~0 solitary particle instantaneously has a sufficiently
slow speed (i.e., $|\vp|\ll mc$), to this particle's nonrelativistic
classical Hamiltonian $\Hnz$ (which corresponds to the quantized Ham%
iltonian operator $\q\Hnz$ of Eq.~(10a)) plus this particle's rest
mass energy $mc^2$.  We therefore regard,
\re{
    P^\mu\eqdf p^\mu + eA^\mu(\vr, t)/c,
}{10b}
as this solitary particle's \e{total} four-momentum.  Eq.~(10b) im%
plies that this particle's relativistic \e{total} three-momentum is,
\re{
    \vP = \vp + e\vA(\vr, t)/c,
}{10c}
and that its relativistic total energy is,
\re{
    E(\vr, \vp, t) = cP^0 = (m^2c^4 + |c\vp|^2)^\hf  + eA^0(\vr, t).
}{10d}
Here we are in the fortunate position of being able to \e{solve}
Eq.~(10c) for the particle's relativistic \e{kinetic} three-momentum
$\vp$ as a \e{function} of its relativistic \e{total} (i.e., \e{canon%
ical}) three-momentum $\vP$ in \e{closed form}, i.e.,
\re{
    \vp(\vP) = \vP - e\vA(\vr, t)/c,
}{10e}
which we must now \e{substitute} into Eq.~(10d) for the relativistic
total energy in order to \e{reexpress} that total energy as the
\e{relativistic Hamiltonian} $\Hrz(\vr, \vP, t)$, i.e.,
\[\Hrz(\vr, \vP, t)\eqdf E(\vr, \vp(\vP), t).\]
With this we obtain from Eqs.~(10d) and (10e) the fully relati%
vistic classical Hamiltonian $\Hrz(\vr, \vP, t)$, that uniquely
corresponds to our original nonrelativistic Hamiltonian operator
$\q\Hnz$ of Eq.~(10a),
\re{
 \Hrz(\vr, \vP, t) = (m^2c^4 + |c\vP - e\vA(\vr, t)|^2)^\hf  + eA^0(\vr, t).
}{10f}
\indent
Because of the \e{presence of the square root} in Eq.~(10f) for
$\Hrz(\vr, \vP, t)$, there could conceivably be an issue regarding
the \e{ordering} of the mutually noncommuting operators $\qvr$ and
$\qvP$ when one attempts \e{quantize} this classical Hamiltonian
$\Hrz(\vr, \vP, t)$ to become the \e{Hamiltonian operator} $\q\Hrz$.
Use of the \e{Hamiltonian phase-space path integral}~\ct{Ka3} with
$\Hrz(\vr, \vP, t)$ in its \e{classical form as given by} Eq.~(10f)
provides one \e{definitive solution} to any such operator-ordering
issue.  Another \e{completely equivalent} solution to this issue
lies with a \e{natural slight strengthening} of Dirac's canonical
commutation rule such that \e{it remains self-consistent}~\ct{Ka4}.
From \e{either} of these approaches the resulting \e{unambiguous
operator-ordering rule} turns out to be the one of Born and Jordan%
~\ct{B-J}.

It is well worth noting that the relativistic classical Hamiltonian
$\Hrz(\vr, \vP, t)$ of Eq.~(10f) for the solitary spin~0 charged
particle, when inserted into Hamilton's classical equations of motion,
yields, after taking Eq.~(10c) into account, \e{the fully relativis%
tic version of the Lorentz-force law}.  In other words, the relativis%
tic solitary charged-particle Hamiltonian $\Hrz(\vr, \vP, t)$ of Eq.~%
(10f) embodies \e{precisely} the well-known classical relativistic
physics of the charged particle developed by H.\ A.\ Lorentz~\ct{L-L}.
We also note that in the limit that the solitary-particle charge $e$
goes to zero, $\Hrz(\vr, \vP, t)$ reduces to the relativistic free-%
particle Hamiltonian $\Hfr$ of Eq.~(1a), as it indeed must.  These re%
sults buttress confidence that the above-described systematic approach
to upgrading physically trustworthy nonrelativistic solitary-particle
Hamiltonians to fully relativistic ones is physically sound.

We now turn to the positive-mass spin~$\hf$ solitary charged particle
in the presence of an external electromagnetic potential $A^{\mu}(\vr,
t)$.  Its nonrelativistic Hamiltonian $\Hnh$ is the \e{same} as the
nonrelativistic Hamiltonian $\Hnz$ of the spin~0 solitary charged par%
ticle \e{except for an additional interaction energy between the ex%
ternal magnetic field and the spin~$\hf$ particle's magnetic dipole
moment due to its intrinsic spin}, i.e., its Pauli spin magnetic di%
pole energy.  Notwithstanding that this Pauli energy is customarily
\e{formally written} as being proportional to $(1/c)$, \e{it must
nonetheless be kept in the nonrelativistic limit} because it \e{fails
to vanish} in the spin~$\hf$ particle's \e{rest frame},
\re{
    \Hnh = |\vp|^2/(2m) + 
           (ge/(mc))(\h /2)\vec\sg\dt(\del_\vr\x\vA(\vr ,t))
           + eA^0(\vr, t).
}{11a}
Just as in the case of $\Hnz$, we \e{deliberately refrain for the time
being} from quantizing $\Hnh$ in its conventional phase-space degrees
of freedom $(\vr, \vp)$ in order to facilitate the derivation of its
natural fully relativistic upgrade.  We cannot, however, switch off
the inherently \e{quantum} nature of the spin~$\hf$ particle's intrin%
sic spin without \e{causing its physical effects to disappear alto%
gether}, so we have \e{no choice} but to accept the Hamiltonian $\Hnh$
of Eq.~(11a) as a \e{two-by-two Hermitian matrix} whose four entries
are (complex-valued) classical dynamical variables.  The Pauli spin
magnetic dipole energy contribution to $\Hnh$ is, however, the \e{only
part} of this nonrelativistic Hamiltonian which is \e{not} a multiple
of the two-by-two \e{identity} matrix.  Now the Lorentz-covariant
four-momenta that we shall be developing in the course of deriving the
natural relativistic upgrade of $\Hnh$ will of course \e{themselves}
naturally come out to be four-vectors of two-by-two matrices, but this
should not present an issue \e{insofar as their four components always
mutually commute}.  To \e{ensure} that this is the case, we shall
``quarantine'' the \e{non-identity} Pauli spin magnetic dipole energy
matrix into a \e{Lorentz scalar}.  We can then render this entity
\e{dimensionless} by dividing it by $mc^2$.  If we now \e{multiply}
this \e{dimensionless Lorentz scalar} by the particle's \e{kinetic}
four-momentum $p^\mu = ((m^2c^2 + |\vp|^2)^\hf, \vp)$, we will indeed
have a Lorentz-covariant four-momentum contribution whose time compo%
nent times $c$ reduces to the Pauli energy matrix in the particle rest
frame, which is precisely what we require.

There remains, of course, the challenging problem of reexpressing the
complicated Pauli energy matrix term of Eq.~(11a) as a Lorentz scalar.
In relativistic tensor language, the magnetic field axial vector
$(\del_\vr\x\vA(\vr, t))$ that appears in the Pauli energy matrix term
of Eq.~(11a) comprises a certain \e{three-dimensional part} of the
four-dimensional relativistic second-rank antisymmetric electromagnet%
ic \e{field tensor} $F^{\mu\nu}(\vr, t) = \pa^\mu A^\nu(\vr, t) -
\pa^\nu A^\mu(\vr, t)$.  Now \e{if} we can manage to reexpress the
spin~$\hf$ angular-momentum axial vector $(\h/2)\vec\sg$ that appears
in the Pauli energy matrix term of Eq.~(11a) as a ``matching'' three-%
dimensional part of a four-dimensional relativistic second-rank anti%
symmetric tensor $s^{\mu\nu}$, hopefully the Pauli energy matrix term
of Eq.~(11a) will end up being proportional to to their \e{Lorentz-%
scalar contraction} $s^{\mu\nu}F_{\mu\nu}(\vr, t)$.  As the first step
toward this goal, we define the natural \e{three-dimensional} second-%
rank antisymmetric spin~$\hf$ tensor $S^{ij}$ in terms of the spin~%
$\hf$ angular momentum axial vector $(\h/2)\vec\sg$,
\[S^{ij}\eqdf (\h /2)\ep^{ijk}\sg^k,\]
and then note that the most complicated factor in the Pauli energy ma%
trix term of Eq.~(11a) neatly reduces to a contraction of $S^{ij}$
with the well-known \e{magnetic-field} three-dimensional part
$F^{ij}(\vr, t)$ of $F^{\mu\nu}(\vr, t)$, i.e.,
\[(\h /2)\vec\sg\dt(\del_\vr\x\vA(\vr ,t)) = (1/2)S^{ij}F^{ij}(\vr, t).\]
This allows us to reexpress the \e{nonrelativistic} Hamiltonian matrix
$\Hnh$ of Eq.~(11a) in the \e{relativistically more suggestive form},
\re{
    mc^2 + \Hnh = mc^2[1  + |\vp|^2/(2m^2c^2) + 
                  (g/2)(e/(m^2c^3))S^{ij}F^{ij}(\vr, t)]
                  + eA^0(\vr, t).
}{11b}
\indent

Of course we need to go \e{beyond} $S^{ij}$ to the spin~$\hf$ parti%
cle's fully covariant \e{four-dimensional} antisymmetric spin tensor
$s^{\mu\nu}$.  In the particle rest frame, namely in the special
inertial frame where the particle kinetic three-momentum $\vp$ vanish%
es, the nine space-space components of $s^{\mu\nu}$ must clearly be
the nine components of $S^{ij}$, and its remaining seven components
must be \e{filled out with zeros}, i.e.,
\[s^{\mu\nu}(\vp = \vz)\eqdf  \dl^\mu_i\dl^\nu_jS^{ij},\]
because this \e{ensures} that, in the particle rest frame,
\[s^{\mu\nu}(\vp = \vz)F_{\mu\nu}(\vr, t) = S^{ij}F^{ij}(\vr, t).\]
Once a tensor is \e{fully determined} in \e{one} inertial frame, it is
fully determined in \e{all} inertial frames by application of the appro%
priate Lorentz transformation to its indices.  To get from the particle
rest frame to the inertial frame where the particle has kinetic three-%
momentum $\vp$ simply requires the appropriate Lorentz-boost four-dimen%
sional matrix $\Lambda^\mu_\al(\vv(\vp)/c)$ that is characterised by the
corresponding dimensionless scaled relativistic particle velocity,
\[\vv(\vp)/c = (\vp/(mc))/(1 + |\vp/(mc)|^2)^\hf,\]
and its accompanying dimensionless time-dilation factor,
\[\gm(\vp) = (1 + |\vp/(mc)|^2)^\hf,\]
so that, in general,
\[s^{\mu\nu}(\vp) = \Lambda^\mu_i(\vv(\vp)/c)
                    \Lambda^\nu_j(\vv(\vp)/c)S^{ij},\]
which, of course, \e{ensures} that $s^{\mu\nu}(\vp)F_{\mu\nu}(\vr, t)$
is a Lorentz scalar that Lorentz-invariantly \e{conveys} the spin~%
$\hf$ particle's \e{rest-frame value of} $S^{ij}F^{ij}(\vr, t)$.

With that, we are in the position to be able to write down the Lo%
rentz-covariant \e{total} four-momentum matrix $P^\mu$ for the spin~%
$\hf$ particle in the presence of the external electromagnetic poten%
tial $A^\mu(\vr, t)$ that corresponds to its nonrelativistic energy
matrix of Eq.~(11b) in the same way that the Lorentz-covariant total
four-momentum $P^\mu$ of Eq.~(10b) for the spin~0 particle in the pre%
sence of $A^\mu(\vr, t)$ corresponds to \e{its} nonrelativistic energy
$(mc^2 + \Hnz)$,
\re{
    P^\mu\eqdf p^\mu[1 + (g/2)(e/(m^2c^3))s^{\al\bt}(\vp)F_{\al\bt}(\vr, t)]
               + eA^\mu(\vr, t)/c.
}{11c}
From $P^\mu$ we obtain the spin~$\hf$ particle's relativistic total
energy matrix,
\re{
E(\vr ,\vp ,t) = cP^0 = (m^2c^4 +|c\vp |^2)^\hf
[1 + (g/2)(e/(m^2c^3))s^{\mu\nu}(\vp )F_{\mu\nu}(\vr ,t)]
+ eA^0(\vr ,t),
}{11d}
and also its relativistic total (i.e., \e{canonical}) three-momentum
matrix,
\re{
\vP = \vp [1 + (g/2)(e/(m^2c^3))s^{\mu\nu}(\vp )F_{\mu\nu}(\vr ,t)]
+ e\vA (\vr ,t)/c.
}{11e}
It is apparent from Eq.~(11e) that we \e{cannot solve} for the spin~%
$\hf$ particle's kinetic momentum matrix $\vp(\vP)$ as a \e{function}
of its \e{canonical} momentum matrix $\vP$ in \e{closed form}, but we
\e{can} express $\vp(\vP)$ in the ``iteration-ready'' form,
\re{
\vp(\vP) = (\vP - e\vA (\vr ,t)/c)
[1 + (g/2)(e/(m^2c^3))s^{\mu\nu}(\vp (\vP ))F_{\mu\nu}(\vr ,t)]^{-1}.
}{11f}
Furthermore, the spin~$\hf$ particle's relativistic total energy ma%
trix $E(\vr, \vp, t)$ of Eq.~(11d) yields the \e{schematic form} of
its \e{relativistic Hamiltonian matrix} $\Hrh(\vr, \vP, t)$ as simply
$E(\vr, \vp(\vP), t)$,
\re{
\Hrh(\vr ,\vP ,t) =
(m^2c^4 + |c\vp(\vP)|^2)^\hf
[1 + (g/2)(e/(m^2c^3))s^{\mu\nu}(\vp (\vP ))F_{\mu\nu}(\vr ,t)]
+ eA^0(\vr ,t).
}{11g}
If we take the limit $g\rta 0$ in Eqs.~(11f) and (11g), then
$\Hrh(\vr ,\vP ,t)\rta\Hrz(\vr ,\vP ,t)$, as is easily checked
from Eq.~(10f).  Of course it is nothing more than 
basic common sense that fully relativistic spin~$\hf$ theory
must reduce to fully relativistic spin~0 theory when the
spin coupling of the single particle to the external field is
switched off, but analogous cross-checking between the Dirac
and Klein-Gordon theories is never so much as discussed!  It
is certainly possible to add a term to the Dirac Hamiltonian
that \e{cancels out} it's \e{supposed} $g = 2$ spin coupling to
the magnetic field, but the result of doing this bears \e{very
little resemblance} to the Klein-Gordon equation in the presence
of the external electromagnetic potential!  Elementary consis%
tency checks are obviously \e{not} the strong suit of those two
``theories''!  If we similarly take the limit $e\rta 0$ in
Eqs.~(11f) and (11g), then $\Hrh(\vr ,\vP ,t)\rta(m^2c^4 +
|c\vP|^2)^\hf$, the free-particle Hamiltonian of Eq.~(1a), as
is physically required.

It is unfortunate that Eq.~(11f) for $\vp(\vP)$ is not amenable
to closed-form solution, but if we assume that the spin coupling term,
$(g/2)(e/(m^2c^3))s^{\mu\nu}(\vp (\vP ))F_{\mu\nu}(\vr ,t)$, which is
a dimensionless Hermitian two-by-two matrix, effectively has the
magnitudes of both of its eigenvalues much smaller than unity (which
should be a very safe assumption for atomic physics), then we can
approximate $\vp (\vP )$ via successive iterations of Eq.~(11f), which
produces the approximation $(\vP - e\vA (\vr ,t)/c)$ for $\vp (\vP )$
through zeroth order in the spin coupling and,
\[\vp (\vP )\approx  (\vP - e\vA (\vr ,t)/c)
[1 + (g/2)(e/(m^2c^3))s^{\mu\nu}(\vP - e\vA (\vr ,t)/c)
F_{\mu\nu}(\vr ,t)]^{-1},\]
through first order in the spin coupling.  We wish to
interject at this point that since $s^{\mu\nu}(\vp (\vP ))$
is an antisymmetric tensor, the tensor contraction
$s^{\mu\nu}(\vp (\vP ))F_{\mu\nu}(\vr ,t)$ is equal to
$2s^{\mu\nu}(\vp (\vP ))\partial_{\mu}A_{\nu}(\vr ,t)$,
which is often a more transparent form. Now if we simply
use the approximation $(\vP - e\vA (\vr ,t)/c)$
through zeroth order in the spin coupling for the kinetic
three-momentum matrix $\vp (\vP )$ of Eq.~(11f), we obtain
the following approximation to the spin~$\hf$ relativistic
Hamiltonian matrix $\Hrh(\vr, \vP, t)$ of Eq.~(11g),
\re{
\Hrh(\vr ,\vP ,t)\approx
(m^2c^4 +|c\vP - e\vA (\vr ,t)|^2)^\hf
[1 + (ge/(m^2c^3))s^{\mu\nu}(\vP - e\vA (\vr ,t)/c)
\partial_{\mu}A_{\nu}(\vr ,t)] + eA^0(\vr ,t).
}{11h}
The \e{approximation} on the right-hand side of Eq.~(11h) to the Ham%
iltonian matrix $\Hrh(\vr ,\vP ,t)$ (whose  schematic form is given
by Eq.~(11g)) for the positive-mass spin~$\hf$ charged relativistic
solitary particle in the presence of the external electromagnetic po%
tential $A^\mu(\vr ,t)$, is a two-by-two matrix whose four entries are
(complex-valued) \e{classical dynamical variables}.  These four en%
tries must each be \e{quantized} in accordance with \e{the Born-Jordan
operator-ordering rule}, analogously to the case of the spin~0 relati%
vistic solitary-particle Hamiltonian $\Hrz(\vr, \vP, t)$ of Eq.~(10f).
Of course higher-order approximations in the spin coupling to the
spin~$\hf$ solitary-particle Hamiltonian matrix $\Hrh(\vr ,\vP ,t)$
must \e{likewise} be quantized.

\subsection*{Antiparticles from field-theory symmetry instead of from
             negative energy}

Let us denote the just-mentioned Born-Jordan \e{quantization} of the
$ij$ entry ($i,j=1,2$) of the Hamiltonian matrix $\Hrh(\vr ,\vP ,t)$
(given schematically by Eq.~(11g)) for the relativistic spin~$\hf$
solitary particle of charge $e$ and positive mass $m$ in the presence
of the external electromagnetic potential $A^\mu(\vr ,t)$ as $\bigl(
\qHrh(e, m, [A^\mu])\bigr)_{ij}$.  Then a basic quantum field-theory
model for \e{electrons alone}, which all have charge $-e$ and mass
$m_-$, in the presence of the external electromagnetic potential
$A^\mu(\vr, t)$ is given by the Hamiltonian operator,
\re{
 \qHF^{(-)} = \isis\psmdi\la\vr|
  \bigl(\qHrh(-e, m_-, [A^\mu])\bigr)_{ij}|\vr'\ra\psmj.
}{12a}
Since the relativistic solitary-particle Hamiltonian $\Hrh(\vr, \vP,
 t)$ has square roots whose arguments involve the canonical momentum
$\vP$, the above-utilized \e{configuration representation} of its
\e{quantization} $\qHrh(e, m, [A^\mu])$ will be \e{nonlocal}, and
therefore the quantum field-theory Hamiltonian operator $\qHF^{(-)}$
clearly \e{cannot} be expressed in terms of a local Hamiltonian
\e{density} in the \e{configuration regime} utilized in Eq.~(12a).

Now a quantum field-theory model which involves electrons \e{alone} is
obviously extremely \e{charge asymmetric}.  To \e{extend} our basic
quantum field-theory model to one which manifests the symmetry of
charge conjugation invariance, we are \e{compelled} to postulate the
existence of \e{another particle} that has the \e{opposite charge} to
that of the electron, but is \e{otherwise identical in all respects to
the electron}.  Denoting the creation fields of this postulated
\e{positron} as $\pspdi$, we readily write down a minimally extended
basic quantum field-theory Hamiltonian operator that indeed manifests
charge conjugation invariance,
\re{\ba{c}
 \qHF^{(-+)} = \isis\bigl[ \\
\m{} \\
  \psmdi\la\vr|
  \bigl(\qHrh(-e, m_-, [A^\mu])\bigr)_{ij}|\vr'\ra\psmj + \\
\m{} \\
  \pspdi\la\vr|
  \bigl(\qHrh(+e, m_-, [A^\mu])\bigr)_{ij}|\vr'\ra\pspj\bigr].
  \ea
}{12b}
\vspace{-0.6\baselineskip}
\m{}\\
\indent
The minimally extended basic quantum field-theory Hamiltonian operator
$\qHF^{(-+)}$ of Eq.~(12b) describes the \e{scattering} (or \e{bind%
ing}) of both relativistic electrons and relativistic positrons by the
external electromagnetic potential $A^\mu(\vr, t)$.  We know, however,
that in principle such a potential could, if it were sufficiently ra%
pidly-varying and strong, produce (or annihilate) electron-positron
\e{pairs}.  We can open a theoretical door to the occurrence of these
electron-positron \e{pair processes} by imposing a \e{further charge-%
related symmetry} on the quantum field-theory Hamiltonian of Eq.~%
(12b), namely its invariance under interchange of electron annihila%
tion with positron creation and also under interchange of positron
annihilation with electron creation.  The simplest extension of
$\qHF^{(-+)}$ which manifests this ``charge equivalence'' symmetry
under the interchanges $\psmi\lra\pspdi$ and $\pspi\lra\psmdi$, and
which \e{as well} maintains the charge conjugation invariance symme%
try, is given by the Hamiltonian operator,
\re{\ba{c}
 \qHF^{(-\lra+^\dg)} = \hf\isis\bigl[ \\
\m{} \\
  (\psmi+\pspdi)^\dg\la\vr|
  \bigl(\qHrh(-e, m_-, [A^\mu])\bigr)_{ij}|\vr'\ra(\psmj+\pspdj) + \\
\m{} \\
  (\pspi+\psmdi)^\dg\la\vr|
  \bigl(\qHrh(+e, m_-, [A^\mu])\bigr)_{ij}|\vr'\ra(\pspj+\psmdj)\bigr].
  \ea
}{12c}
\vspace{-0.6\baselineskip}
\m{}\\
It is apparent from Eq.~(12c) that the imposition of the twin symme%
tries of charge conjugation invariance and ``charge equivalence'' does
indeed produce a quantum field-theory model for electron-positron pair
creation and annihilation by the external electromagnetic potential
$A^\mu(\vr, t)$, as well as electron and positron scattering (or bind%
ing) by that potential.

Now the visible universe is obviously skewed toward the preponderance
of electrons over positrons, so we certainly expect that there is a
physical agency which \e{breaks} charge conjugation invariance.  While
there is experimental evidence of such an agency, contemporary theo%
retical physics has not yet understood it in more than phenomenologi%
cal fashion, but one would suppose that there must exist fields whose
effective interaction strength magnitudes with electron and positron
are \e{unequal}.  There is, of course, \e{no} apparent theoretical
reason why such a charge conjugation invariance \e{breaking} mechanism
shouldn't generate \e{disparate corrections} to the electron and posi%
tron \e{masses}.  In fact, from the ``laziest'' phenomenological
standpoint, the \e{simplest} way to introduce charge conjugation in%
variance \e{breaking} into our model field-theory Hamiltonian operator
$\qHF^{(-\lra+^\dg)}$ of Eq.~(12c) is to insert into it \e{exactly}
such a \e{mass difference} $\dl m$ between positron and electron,
\re{\ba{c}
 \qHF^{(-\lra+^\dg)_{\mr{broken}}}(\dl m) = \hf\isis\bigl[ \\
\m{} \\
  (\psmi+\pspdi)^\dg\la\vr|
  \bigl(\qHrh(-e, m_-, [A^\mu])\bigr)_{ij}|\vr'\ra(\psmj+\pspdj) + \\
\m{} \\
 (\pspi+\psmdi)^\dg\la\vr|
 \bigl(\qHrh(+e, m_-+\dl m, [A^\mu])\bigr)_{ij}|\vr'\ra(\pspj+\psmdj)\bigr].
 \ea
}{12d}
\vspace{-0.6\baselineskip}
\m{}\\
This simple-minded model of charge conjugation invariance breaking
drives home the point that in the \e{absence} of the theoretically
ill-founded Klein-Gordon and Dirac equations---with their unphysical
free particles of unboundedly negative energies (amongst a \e{pletho%
ra} of other unphysical features), upon which is further erected, via
the mind-boggling assumption of compulsory negative-energy free-parti%
cle travel backwards in time, the ``species identity'' of antiparticle
with particle which implies their perfect mass equality and the CPT
theorem---there is \e{simply no compelling reason whatsoever} to ex%
pect the \e{exact equality} of particle and antiparticle masses.  On
the contrary, it would be \e{entirely unexpected} for the breaking of
particle-antiparticle symmetry to fail to \e{naturally split} particle
and antiparticle masses.  Thus the embrace of the correspondence prin%
ciple in relativistic quantum theory \e{removes} that discipline's
categorical incompatibility with the preliminary MINOS finding of a
mass difference between muon antineutrino and neutrino~\ct{Vh, Cw}---%
as well as removing the CPT theorem and the configuration-regime
\e{locality} of relativistic quantum field theory.

The relativistic quantum electrodynamics implied by embrace of the
correspondence principle clearly differs in detail from the ``ortho%
dox'' discipline bearing that name, with the pervasive influence of
the ``minimally coupled'' Dirac Hamiltonian operator obviously sup%
planted by the Hamiltonian operators $\qHrh(\mp e, m_-, [A^\mu])$
that feature in Eq.~(12c) above.  The disagreement between the elec%
tronic hydrogen and muonic hydrogen approaches to measuring the charge
radius of the proton~\ct{Po}, with their differing degrees of depend%
ence on calculated quantum electrodynamics contributions, might turn
out to be a harbinger of the need to reposition relativistic quantum
electrodynamics firmly on the foundation of the correspondence
principle.

\subsection*{Conclusion}

An immense amount of work will need to be carried out in order to give
birth to a comprehensive relativistic quantum electrodynamics (or oth%
er relativistic quantum field theory) that is properly founded on the
correspondence principle.  The proximate task is to upgrade the model
field-theory Hamiltonian operator $\qHF^{(-\lra+^\dg)}$ of Eq.~(12c)
to accommodate the \e{quantized} electromagnetic potential.  This is a
tricky undertaking: because electromagnetism is a gauge theory, only a
\e{part} of it is dynamical and quantizable, but its nondynamical,
nonquantizable potential part still has \e{physical consequences},
while \e{relativistically compatible gauge fixing} is needed to block
spurious \e{unphysical} consequences~\ct{Ka5}.  After electromagnetism
has been (hopefully) successfully dealt with, the ``Feynman rules''
threaten to be a tangled web indeed: the $\Hrh(\vr, \vP, t)$ schemati%
cally given by Eq.~(11g) can itself only be obtained \e{iteratively},
and, even that aside, its square root structure, taken in conjunction
with Eq.~(11f), \e{already} guarantees that it depends on the electron
charge $e_-$ to arbitrarily high order.  To this must be added the re%
quirement of its Born-Jordan quantization to obtain $\qHrh(e, m,
[A^\mu])$, on \e{top} of which comes perturbative development of the
consequent quantum field theory in order to calculate transition am%
plitudes!  There can be no question that sustained, patient and ingen%
ious efforts by many contributors over a very extended period of time
will be essential to obtaining results from relativistic quantum elec%
trodynamics founded on the correspondence principle.


\begin{thebibliography}{15}
\bibitem{Vh}
P. Vahle for the MINOS Collaboration,
Presentation at the XXIV International Conference on Neutrino Physics
and Astrophysics (Neutrino 2010) in Athens, Greece on June 14, 2010.
\bibitem{Cw}
R. Cowen,
ScienceNews \textbf{178}, No.\ 2,
9 (2010).
\bibitem{Ka0}
S. K. Kauffmann,
arXiv:1005.2641 [physics.gen-ph]
(2010).
\bibitem{Po}
R. Pohl, et al.,
Nature \textbf{466},
213 (2010).
\bibitem{Scf}
L. I. Schiff,
\e{Quantum Mechanics}
(McGraw-Hill, New York, 1955).
\bibitem{B-D}
J. D. Bjorken and S. D. Drell,
\e{Relativistic Quantum Mechanics}
(McGraw-Hill, New York, 1964).
\bibitem{Dir}
P. A. M. Dirac,
\e{The Principles of Quantum Mechanics}
(Oxford University Press, London, 1958).
\bibitem{D28}
P. A. M. Dirac,
Proc.\ Roy.\ Soc.\ (London) \textbf{A117},
610 (1928).
\bibitem{Scw}
S. S. Schweber,
\e{An Introduction to Relativistic Quantum Field Theory}
(Harper \& Row, New York, 1961).
\bibitem{B-S}
N. N. Bogoliubov and D. V. Shirkov,
\e{Introduction to the Theory of Quantized Fields}
(Interscience Publishers, New York, 1959).
\bibitem{Ka1}
S. K. Kauffmann,
arXiv:0909.4025 [physics.gen-ph]
(2009).
\bibitem{Ka3}
S. K. Kauffmann,
arXiv:0910.2490 [physics.gen-ph]
(2009).
\bibitem{Ka4}
S. K. Kauffmann,
arXiv:0908.3755 [quant-ph]
(2009).
\bibitem{B-J}
M. Born and P. Jordan,
Z.\ Physik \textbf{34},
858 (1925).
\bibitem{L-L}
L. D. Landau and E. M. Lifshitz,
\e{The Classical Theory of Fields}
(Butterworth-Heinemann, Oxford, 1975).
\bibitem{Ka5}
S. K. Kauffmann,
arXiv:1005.1101 [physics.gen-ph]
(2010).
\end{thebibliography}
\end{document}